# Room temperature biaxial magnetic anisotropy in La$_{0.67}$Sr$_{0.33}$MnO$_3$ thin films on SrTiO$_3$ buffered MgO (001) substrates for spintronic applications

Sandeep Kumar Chaluvadi,[1] Fernando Ajejas,[2,6] Pasquale Orgiani,[3,4] Olivier Rousseau,[1] Giovanni Vinai,[3] Aleksandr Yu Petrov,[3] Piero Torelli,[3] Alain Pautrat,[5] Julio Camarero,[2,6] Paolo Perna,[2] and Laurence Mechin[1,a)]

[1]Normandie Univ, UNICAEN, ENSICAEN, CNRS, GREYC, 14000 Caen, France

[2]IMDEA-Nanociencia, Campus de Cantoblanco, 28049 Madrid, Spain

[3]CNR-IOM TASC National Laboratory, Area Science Park- Basovizza, 34149 Trieste, Italy

[4]CNR-SPIN, UOS Salerno, 84084 Fisciano (SA), Italy

[5]Normandie Univ, UNICAEN, ENSICAEN, CNRS, CRISMAT, 14050 Caen, France

[6]Universidad Autónoma de Madrid, Campus de Cantoblanco, 28049 Madrid, Spain

Spintronics exploits the magnetoresistance effects to store or sense the magnetic information. Since the magnetoresistance strictly depends on the magnetic anisotropy of the system, it is fundamental to set a defined anisotropy to the system. Here, we investigate by means of vectorial Magneto-Optical Kerr Magnetometry (v-MOKE), half-metallic La$_{0.67}$Sr$_{0.33}$MnO$_3$ (LSMO) thin films that exhibit at room temperature pure biaxial magnetic anisotropy if grown onto MgO (001) substrate with a thin SrTiO$_3$ (STO) buffer. In this way, we can avoid unwanted uniaxial magnetic anisotropy contributions that may be detrimental for specific applications. The detailed study of the angular evolution of the magnetization reversal pathways, critical fields (coercivity and switching) allows for disclosing the origin of the magnetic anisotropy, which is magnetocrystalline in nature and shows four-fold symmetry at any temperature.

Half-metallic perovskite oxides promise great advantages over conventional spintronics metallic materials for applications such as magnetic sensors, magnetic random access memory (MRAM), magnetic tunnel junctions (MTJs) and domain wall race-track memories.[1] Perovskite oxides, in general, appear to be a new contender for many novel applications that were considered traditionally beyond its range.[2] The conduction mechanism in these materials, in fact, strongly depends on the interplay between orbital and spin degrees of freedom[3] that may be exploited to add multiferroic or ferroelectric functionalities.[4–6] The complexity of such mechanism, however, can determine entangled magneto transport response becoming in some cases undesired. For example, it has been seen that a switchable anisotropic magnetoresistance (AMR) response in manganites may be hidden by the colossal magnetoresistance (CMR) if the magnetic anisotropy of the system is not accurately designed.[7] Irrespective of the applications, one of

---

[a)] **Author to whom correspondence should be addressed:** laurence.mechin@ensicaen.fr

the key properties that need to be considered for a ferromagnetic sample is the magnetic anisotropy that dictates the magnetization reversals pathways as well as the MR output.[8] While a defined *uniaxial anisotropy* is essential for magnetic field sensors based on AMR,[7] a *biaxial anisotropy*, which provides four stable magnetization states, has capability to encode more information (four binary bits: "00", "01", "10", "11"), and can be used in memory and logic devices.[9,10] However, especially in half-metallic perovskite compounds with four-fold magnetocrystalline anisotropy in bulk, it results difficult to avoid either strain (induced by the substrate) or shape uniaxial anisotropy in thin films.[11,12]

Among manganites, $La_{0.67}Sr_{0.33}MnO_3$ (LSMO) has arisen special interest for its peculiar properties such as nearly 100% spin polarization and room temperature (RT) ferromagnetism with Curie temperature of about $T_c$~370 K, hence, enabling RT spintronic applications. The magnetotransport properties of these compounds in thin films are strongly affected by external perturbations such as substrate-induced strain.[13] In general, in tensile (compressive) strained films, the electron occupancy in '$e_g$' doublet favors in-plane (out-of-plane) $x^2-y^2$ ($3z^2-r^2$) orbitals[14] determining in-plane (out-of-plane) magnetization easy axis.[15] In addition, tensile (compressive) strain reduces (enhances) the Tc with respect to the bulk value.[16,17] Therefore, the choice of the substrate for the LSMO growth is extremely important. The most commonly used single crystal substrate is $SrTiO_3$ (STO) ($a_{STO}$=0.3905 nm) with (001) crystallographic orientation since LSMO ($a_{LSMO}$= 0.387 nm) grown epitaxially on STO exhibits low tensile strain (0.82%) and with good structural and morphological properties.[18] However, in the low thickness regime, a sizeable uniaxial (two-fold) anisotropy contribution[19] due to surface steps and terraces originated from the mis-cut angle of the substrate[7,11,20,21] generally hide the biaxial (four-fold) magnetocrystalline anisotropy of the LSMO film. In the higher thickness regime, LSMO on STO (001) usually show a competition between (biaxial) bulk[12] and shape or strain (uniaxial) magnetic anisotropy.[22] Another single crystal oxide that can be used as a substrate for the growth of LSMO films is MgO with (001) crystallographic orientation. However, the lattice mismatch between LSMO and MgO ($a_{MgO}$= 0.4212) results in large tensile strain (~8%). This generally degrades the film magneto-transport and morphological properties that limited its use.

One of the utmost challenges is to control the magnetic anisotropy of the system via substrate-induced strain and surface engineering. In this letter, to define magnetic anisotropy symmetry in LSMO thin films, instead of growing it directly on cubic STO (001) crystals; we deposited 50 nm LSMO film on cubic MgO (001) substrate and employed 12 nm STO as a buffer layer. The inclusion of 12 nm STO buffer layer on MgO substrate not only helps in incorporating the structural and interfacial defects in it but also minimizes the misfit strain and acts as a template for depositing good quality epitaxial LSMO film. The angular dependent magnetic anisotropy symmetry in our STO buffered LSMO films shows dominant biaxial anisotropy even at RT, enhanced by an order of magnitude at 40 K.

The STO (001) buffered LSMO thin films were epitaxially grown on MgO (001) substrate by pulsed laser deposition (PLD) from commercial stoichiometric targets by using KrF excimer laser of wavelength 248 nm. High energetic laser pulses (1.4 – 1.7 J/cm$^2$) were used for transferring correct elemental ratio of heavy elements for growing stoichiometric LSMO and STO films.[23,24] The deposition was made at 0.35 mbar oxygen pressure while maintaining the substrate temperature at 720 °C. After deposition, the samples were cooled down to RT at 10 °C per minute in 7x10$^2$ mbar oxygen pressure. The thicknesses of LSMO and STO layers were set at 50 and 12 nm, respectively. The structural, morphology, magnetic and electrical transport measurements were done by PANalytical X'Pert four-circle X-Ray Diffraction (XRD) in low and high-resolution modes, Atomic Force Microscopy (AFM), Superconducting Quantum Interference Device (SQUID), and four-probe technique, respectively. Angular dependent *in-plane* magnetization reversal process, coercivity, and magnetic anisotropy measurements were performed at 300 K and 40 K by using vectorial Magneto-Optical Kerr (v-MOKE) magnetometry.[25]

The overall optimal structural and compositional qualities of the film are confirmed by the transport and morphological properties. In fact, resistivity and magnetization vs. temperature measurements (see Fig. S1 in the Supporting Information) demonstrate low residual resistivity, Metal-Insulator transition temperature '$T_{MI}$' and Curie temperature '$T_C$' close to the bulk values, i.e. >420 and 362 K, respectively. Whereas, the $T_{MI}$ of LSMO film, when directly grown on MgO (001) showed reduced value ($T_{MI}$ ~325 K) (see Fig. S2 Supporting Information). Therefore, STO buffer layer helps to reduce structural defects in LSMO layer and improves the film properties. The film surface probed by AFM reveals an RMS (root mean square) roughness of about 0.35 nm (very smooth, i.e. less than one unit cell). The epitaxial structure of the STO and LSMO films demonstrated by the θ-2θ XRD spectrum that shows only (00l) peaks, indicating the preferential c-axis orientation along the [00l] substrate crystallographic direction (Fig. 1(a)). The calculated 'c' lattice parameters of STO and LSMO films are 0.3903 and 0.3873 nm, respectively, that indirectly indicates the optimal oxygen composition in the film.[26] In order to determine the in-plane epitaxial relationship between the LSMO film and MgO (001) substrate, azimuthal ϕ-scans performed around the (0-24) MgO and (0-13) LSMO asymmetric reflections, as shown in Fig. 1(b). The peaks with separation of 90 degrees observed for both MgO substrate and LSMO film demonstrates the four-fold symmetry with the [100] plane of the film parallel to [100] plane of the substrate.

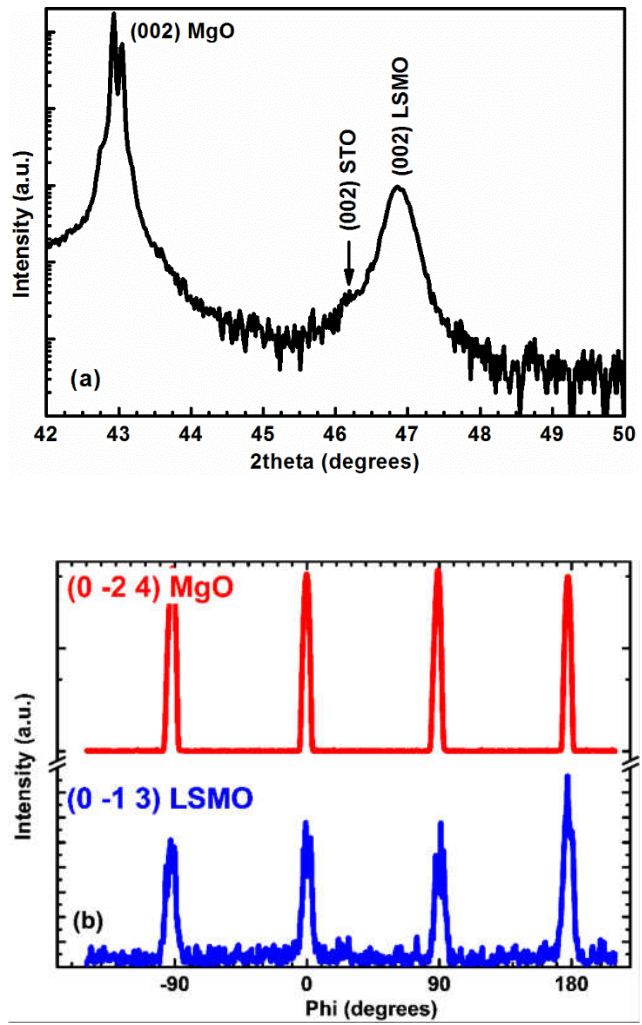

**FIG. 1: (a) Theta-2 theta XRD scan of 50 nm thick LSMO film on STO buffered MgO (001) substrate, (b) asymmetrical Phi scans around (0-13) peak of LSMO (001) film and (0-24) peak of MgO (001) substrate shows 90° separation cube-on-cube epitaxy.**

In order to investigate in details the LSMO film structure, we have performed XRD reciprocal space map (RSM) around the asymmetric (0-13) LSMO and (0-24) MgO reflections. Interestingly, in low-resolution mode (Fig. 2(a)), a single diffraction peak is evident thus inferring a cubic structure of the film (i.e., identical in-plane lattice parameters, a = b). However, in high-resolution mode (i.e., using a Ge (220) double-bounce monochromator), the RSM around the (0-13) LSMO crystallographic reflection shows a double-peaks structure along the Qx in-plane direction (inset of Fig. 2(a)). Such a feature indicates the presence of two slightly different in-plane lattice parameters (i.e. a ≠ b) thus supporting a possible orthorhombic or monoclinic structure for the LSMO film. The average in-plane lattice parameter of LSMO is 0.388 nm, which is very close to the relaxed pseudocubic lattice value. Fig. 2(b) shows the omega scans measured around the LSMO (002) peak at different $\phi$ angles (0°, 45° and 90°). At $\phi = 0°$ and 90°, the presence of double-peak along the LSMO out-of-plane direction indicates, as a first approximation, the presence of two LSMO domains with the c-axis direction tilted of ~0.4° with respect to the [001]-MgO direction. However, rocking curve taken at $\phi = 45°$, does not show any double-peak structure. Note that in phi-scan in Fig.1b, we do have a 6° broad peak, which might "cover-up" the two peaks expected for a rhombohedral structure, and the observed 0.4° splitting of the rocking curves in Fig.2b might be correlated to the actual "rhombohedral" arrangement of the LSMO unit cells. Therefore, from the phi scans ($\alpha=\beta=90°$) and omega scans ($\gamma\neq90°$), we deduce that the most probable scenario is based on a 4-monoclinic domain structure.

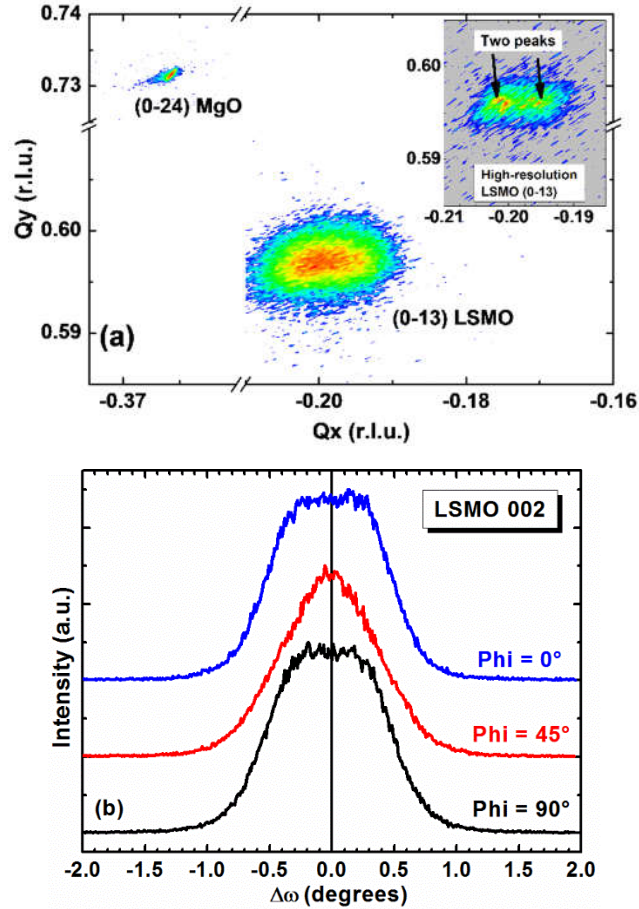

FIG. 2: (Color online) (a) Low-resolution RSM scans of (0-13) LSMO and (0-24) MgO peaks shows that the LSMO film is fully relaxed. The inset presents the high-resolution RSM scan of (0-13) LSMO that shows two peaks, marked with arrows. (b) Omega scans measured around LSMO (002) peak at different phi angles i.e., 0°, 45° and 90°.

The magnetic anisotropy of the film also reveals a four-fold symmetry. This was investigated at RT in details by acquiring *in-plane* Kerr hysteresis loops by v-MOKE in the full angular range (i.e., 0°-360°) while keeping fixed the external magnetic field direction. At θ=0°, the applied external magnetic field is aligned parallel to the [100] crystallographic axis of MgO (001) substrate. By exploiting the simultaneous acquisition of two in-plane magnetization components, i.e., parallel ($M_\parallel(H)$) and perpendicular ($M_\perp(H)$), we can deduce the symmetry of magnetic anisotropy present in the film.[7,11,27] In fact, by inspecting the change of sign in the $M_\perp(H)$, we can accurately locate the easy axis (e.a.) and hard axis (h.a.) directions. Fig. 3 presents the normalized Kerr hysteresis loops of the LSMO (001) film acquired at and around e.a. and h.a. i.e., θ=45°, 90°, and ±9°. Square loop with sharp irreversible transitions in the $M_\parallel(H)$ is found at θ=45°, whereas the $M_\perp(H)$ component is almost zero. This indicates that the magnetization is aligned in the film plane either parallel or anti-parallel, which corresponds to a magnetization reversal proceeding by nucleation and further propagation of magnetic domains (characteristic of an e.a. behavior). When the field is applied away from the e.a., smoother reversible transitions are found indicating that the reversal starts by rotation followed by propagation of domains. In particular, when the magnetic field is applied around the e.a., i.e. from θ=36° to 54°, the magnetization switches with just one irreversible transition and from the $M_\perp(H)$ component, we could observe that there is a significant change of sign. This means that the magnetization rotates within the film plane in clockwise to anticlockwise direction or vice-versa with respect to the applied field angle. The remanence of $M_\parallel(H)$ at the e.a. is $M_{R,\parallel} \cong M_S$ and the coercive field is about 1.8 mT.

The M(H) loops taken at θ=81° and θ=99° shows that the magnetization reversal takes place in three steps, which are marked with arrows in $M_\parallel(H)$ and circles in $M_\perp(H)$ (Fig. 3 (top right)), respectively. Sweeping the filed from positive to negative values, the reversal occurs by a first reversible magnetization rotation, followed by two irreversible transitions that take place by nucleation and propagation of two consecutive 90° domain walls.[28] At h.a. i.e., θ=90°, as the field strength decreases, the $M_\parallel(H)$ loop show rotation of the magnetization followed by sharp irreversible transition and final rotation towards the applied field direction. The reversal proceeds thus with one (two) irreversible transition, related to nucleation and propagation of 180° (90°) domain walls, when the field is applied close to one of the two e.a. (h.a.) orientations of magnetization. These are the typical signatures of a four-fold magnetic symmetry.[27] The experimental data have been properly reproduced in the whole angular range with a modified coherent rotation Stoner-Wohlfarth model[11,27] by using exclusively a single magnetic anisotropy term with four-fold symmetry whose strength was extracted from the experimental curves at the hard axis ($H_K^b$=5 mT). To note, since this model is based on coherent rotation, it fails in reproducing the experimental data in the regions in which nucleation and propagation of domains mechanisms dominate (i.e., close to e.a.). However, it provides a qualitative and quantitative estimation of the relevant anisotropy contributions involved (at the h.a. region).

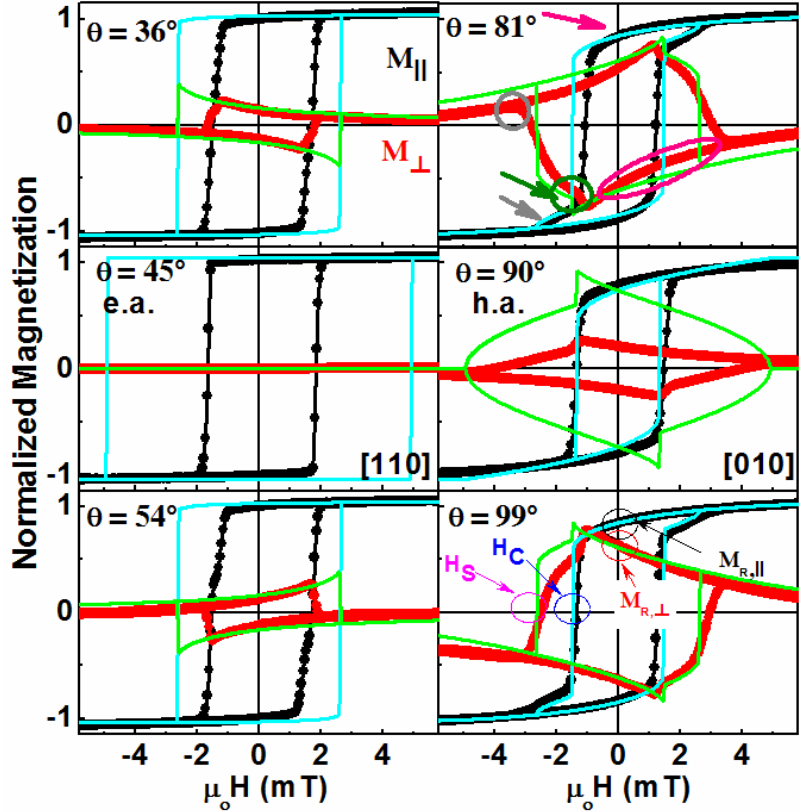

**FIG. 3: (Color online) Normalized magnetic hysteresis cycles of 50 nm thick LSMO film grown on STO buffered MgO (001) measured by v-MOKE at 300 K at and around easy (e.a.) (left) and hard (h.a.) (right) axis. The corresponding applied external magnetic field angles (θ) with respect to the crystallographic axis are specified in the figures. The $M_\parallel$(H) (black) and $M_\perp$(H) (red) loops acquired simultaneously are shown. The arrow (circles) in the top right panel indicates the double transition, which is the signature of biaxial anisotropy. The cyan and green solid lines correspond to simulation results of Stoner-Wohlfarth model.**

The four-fold symmetry of the magnetic anisotropy becomes more evident by plotting the angular dependent remanence and critical fields in the full angular range (Fig. 4). The normalized remanence magnetization plots ($M_{R,\parallel}/M_S$ and $M_{R,\perp}/M_S$) extracted from the experimental $M_\parallel$(H) and $M_\perp$(H) loops at the applied field $\mu_o H=0$ as a function of angle 'θ' are shown in Fig. 4(a). Both the magnetization components show repeating features with the periodicity of 90°. In addition, $M_{R,\perp}$ changes its sign for every 45° i.e., whenever it crosses the characteristic axes. The polar plot of $M_{R,\parallel}/M_S$, and $M_{R,\perp}/M_S$ are presented in panels c-d of Fig. 4 revealing the four-fold symmetry. In particular, the $M_{R,\parallel}/M_S$ resembles a butterfly structure with the highest and lowest values pointing towards e.a. and h.a. of film, i.e., along with [110] and the [010] crystallographic directions, whereas the $M_{R,\perp}/M_S$ shows four lobe symmetrical shape with positive and negative values are depicted in solid and open circles.

Fig. 4(b) shows the angular dependence of the critical fields i.e., coercive ($H_C$) and switching ($H_S$) fields, extracted from $M_\parallel(H)$ and $M_\perp(H)$ loops. The value of '$H_C$' ($H_S$) is higher (lower) at e.a. i.e., [110] and decreases (increases) as it approaches towards h.a. i.e., [010]. The $H_C$ and $H_S$ coincide at and around e.a. corresponds to one irreversible transition (grey shaded area) leading to 180° domain walls reversals. As the field is applied away from the e.a. (i.e., approaching the h.a., white region), $H_S$ increases and reaches the maximum exactly at the h.a. In the white shaded regions, the magnetization reversal takes place with two irreversible transitions that relate to the nucleation and propagation of two consecutive 90° domain walls. The polar plots of the $H_C$ and $H_S$ are presented in Fig. 4(e, f) show symmetrical four lobes and asteroid shape, respectively, with 90° periodicity. From these angular dependent analyses, it is evident that the angles between two adjacent e.a. and h.a. are orthogonal to each other and the minima of $M_{R,\parallel}$ at the consecutive h.a. have identical values, which experimentally prove that no uniaxial magnetic anisotropy contribution exists.[27] These features demonstrate the four-fold symmetry of the magnetic anisotropy in the LSMO/STO/MgO (001) structure, which is similar to the magnetocrystalline anisotropy of the bulk LSMO with easy axes aligned towards 45° [110].[29] The analysis of both magnetization components allows therefore for disclosing with high accuracy the symmetry of the magnetic anisotropy. To note that, even though the authors in the Ref.12 claim to have biaxial anisotropy in LSMO/STO film, the difference between the magnetization remanence states at two consecutive h.a. reveals an additional uniaxial contribution, as explained by the authors in the Ref. 19 and 27.

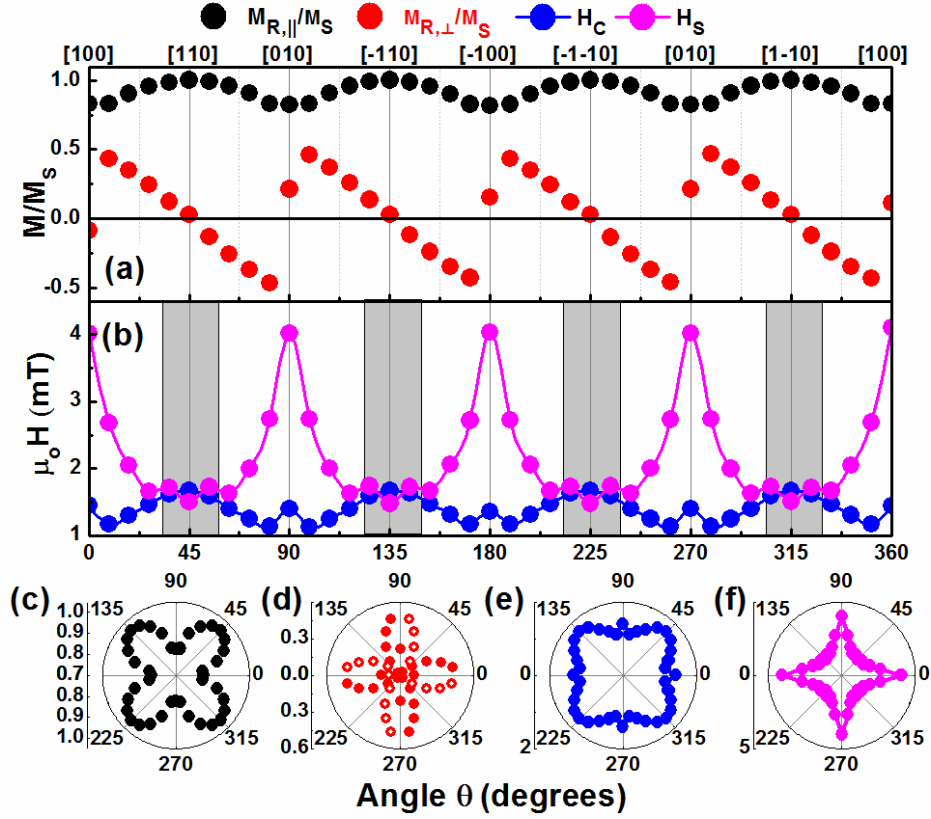

FIG. 4: (Color online) Angular evolution at 300 K of the magnetic properties of 50 nm thick LSMO film grown on STO buffered MgO (001) substrate. (a) Normalized remanence magnetization $M_{R,\parallel}/M_s$ (black) and $M_{R,\perp}/M_s$ (red), (b) critical fields (coercivity '$H_C$' (blue) and switching '$H_S$' (magenta)) as a function of the applied field angle 'θ' shows well-defined 90° periodicity i.e., a pure bi-axial anisotropy. The grey shaded regions in (b) indicate the system exhibits only one irreversible transition, whereas, in the white regions, the system exhibits two consecutive irreversible transitions. (c-f) Polar plots of $M_{R,\parallel}/M_s$, $M_{R,\perp}/M_s$, $H_C$ and $H_S$ respectively. Positive and negative values in (d) represent by solid and open circles.

In order to disclose the nature of the four-fold magnetic anisotropy in our systems, we performed temperature dependent studies. The normalized Kerr hysteresis loops (Fig. 5(a)) measured at 40 K at different in-plane magnetic field directions present similar features and symmetry with respect to the RT case. The e.a. and h.a. are present at 45° [110] and 90° [010]. Near h.a. i.e., at θ=70° and 80°, two irreversible transitions are observed owing to nucleation and propagation of two consecutive 90° domain walls. In this case, the data have been reproduced by using again a single magnetic anisotropy term with four-fold anisotropy and the anisotropy field is one order of magnitude larger than the one found at RT ($H_K^b$=40 mT). Temperature-dependent coercive fields (Fig. 5(b)) in the angular range of 0°-180° has 90° periodicity as the RT behavior but with a tremendous increase of the coercive fields of about one order of magnitude. Therefore, as the temperature decreases from 300 K to 40 K, the signatures of biaxial anisotropy become more evident. The cause of anisotropy is due to magnetocrystalline anisotropy of LSMO, which is usually dominant at low temperatures.[30]

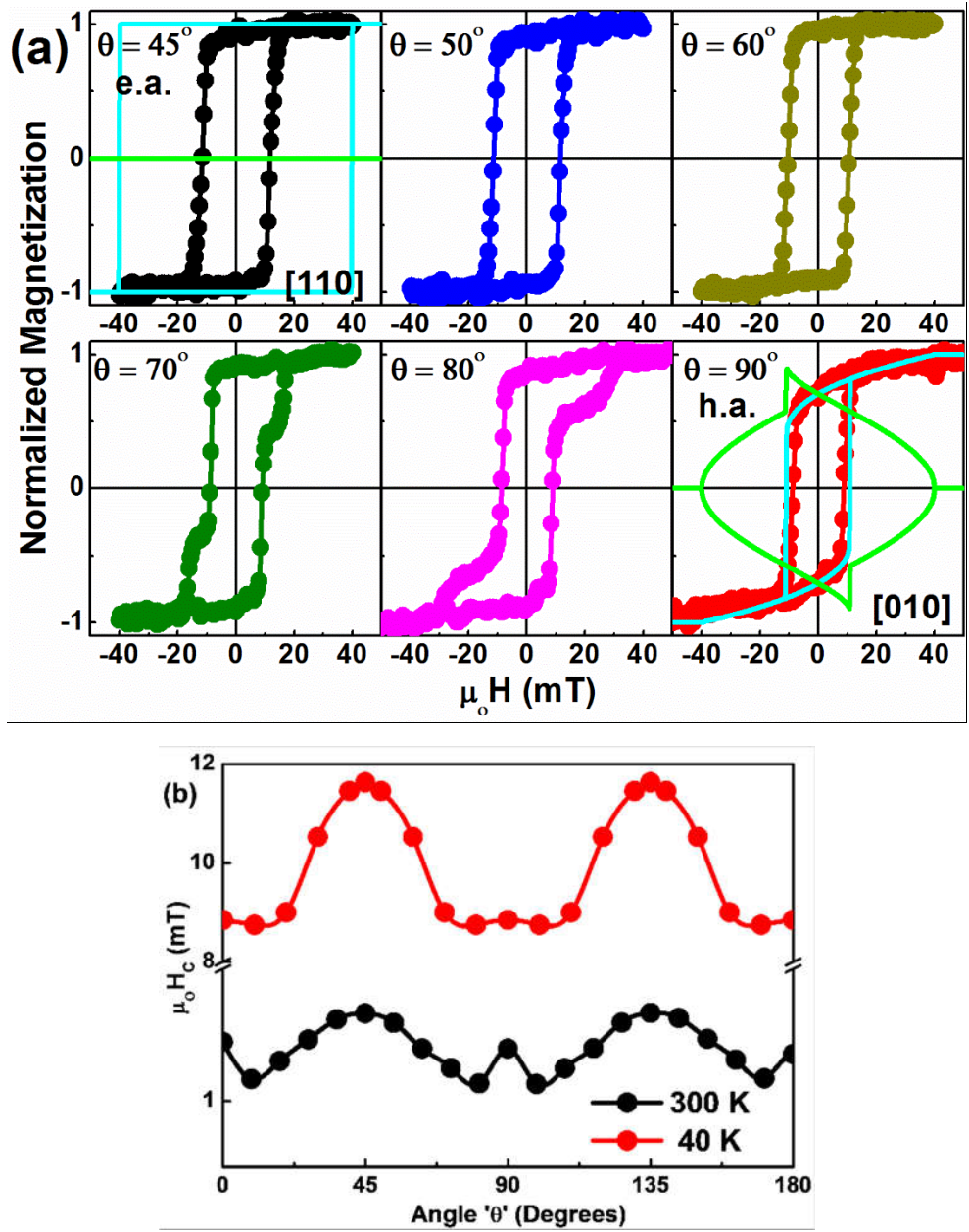

**FIG. 5:** (Color online) (a) Normalized magnetic hysteresis loops of 50 nm thick LSMO film grown on STO buffered MgO (001) around the first half quadrant measured at 40 K, and (b) Angular dependent coercive field '$H_C$' at 40 and 300 K. The cyan and green solid lines in (a) corresponds to simulation results of Stoner-Wohlfarth model.

In summary, we have fabricated high-quality epitaxial LSMO thin films on STO buffered MgO (001) substrate by PLD technique and studied *in-plane* magnetic anisotropy properties at 300 and 40 K. We demonstrated that the use of STO buffer layer improved the quality of LSMO film by accommodating structural defects. The LSMO film has a 4-domain monoclinic structure with domains present along [100] and [010] directions. The magnetic anisotropy also showed biaxial anisotropy with the e.a. direction is aligned towards 45° or [110], where the tilted domains are absent and the h.a. is aligned along the tilted monoclinic domains i.e., 0° or [100]. As temperature decreases, the strength of the biaxial anisotropy increases, being one order of magnitude larger at 40 K, revealing its magnetocrystalline origin. Unlike the LSMO films directly grown on STO (001) substrates that show anomalies in magnetic anisotropy, films that were grown onto STO buffered MgO (001) substrates displays well-defined biaxial anisotropy, that can be useful in four bits logic devices based on purely anisotropic magnetoresistance response.

See supplementary information for morphology, magnetic and electrical transport properties of LSMO/STO/MgO (001) film.

SKC acknowledge Programme International de Coopération Scientifique (PICS) du CNRS under grant no. 6161 and SIMEM Doctoral School at Universite de Caen Normandie for financial support. IMDEA-Nanociencia acknowledges support from the 'Severo Ochoa' Program for Centres of Excellence in R&D (MINECO, Grant SEV-2016-0686). FA, JC, and PP acknowledge the support of Spanish MINECO Projects No. FIS2015-67287-P and FIS2016-78591-C3-1-R, and from the Comunidad de Madrid through Project NANOFRONTMAG CM. This project has received funding from the European Union's Horizon 2020 research and innovation programme under grant agreements No. 737116 (byAxon) and No. 654360 NFFA-Europe.